\documentclass[aps,pra,twocolumn,superscriptaddress,longbibliography]{revtex4-2}
\usepackage{amsmath,amssymb,amsthm,bm,graphicx,mathtools}
\usepackage{enumitem}
\setlist{itemsep=1pt,topsep=3pt,parsep=0pt}
\usepackage[colorlinks=true,linkcolor=blue,citecolor=blue,urlcolor=blue]{hyperref}
\newtheorem{theorem}{Theorem}
\newtheorem{proposition}{Proposition}

\theoremstyle{remark}

\newcommand{\E}{\mathbb{E}}
\newcommand{\Kt}{\tilde K}
\renewcommand{\Pr}{\mathbb{P}}

\begin{document}

\title{Information--control limits for record-indistinguishable reset of a monitored qubit}
\author{Sinan Bugu}
\email{sbugu@ncsu.edu}
\affiliation{Department of Physics and Astronomy, North Carolina State University, Raleigh, North Carolina 27695, USA}
\affiliation{BuQuLab LLC, Winston-Salem, North Carolina, USA}

\author{Fatih Ozaydin}
\affiliation{Tokyo International University, 4-42-31 Higashi-Ikebukuro, Toshima-ku, Tokyo, 170-0013, Japan}
\affiliation{Nanoelectronics Research Center, Kosuyolu Mah., Lambaci Sok., Kosuyolu Sit., No:9E/3 Kadikoy, Istanbul, 34718, Türkiye}

\date{\today}

\begin{abstract}
A qubit that is continuously monitored from the moment it is prepared in its excited state writes evidence of that preparation into its measurement record. We ask how much control is needed to reset such a qubit to its ground state while keeping its complete record statistically close to the record of an undriven ground-state qubit. The qubit is measured in $\sigma_z$ at rate $\Gamma_m$ with efficiency $\eta$ and has no relaxation. For feedback that acts causally on the record, distinguishability is the Kullback--Leibler divergence $D$ of the controlled record law from the reference law, and the control cost is the quadratic effort $J$, the integrated squared Rabi frequency. We prove that every open-loop drive leaves $D$ infinite and that every admissible control obeys $DJ\ge\Gamma_mU(\eta,D/\kappa_0)$, where $U$ is given by a one-parameter family of integrals and $\kappa_0=1-\ln2$. Our main result is the exact high-effort constant $\lim_{\epsilon\downarrow0}\inf\{DJ/\Gamma_m: D\le\epsilon\}=K(\eta)$, where $K(\eta)=[\int_0^\pi\sqrt{2\eta(1+\cos\theta)^2+(1-\eta)\sin^2\theta}\,d\theta]^2$; an explicit sequence of two-phase feedback policies attains it. The first term of the integrand is the population evidence that the record receives while the qubit is rotated toward the ground state. The second is the price of the impurity created when the unrecorded part of the measurement dephases the coherence that the rotation needs; the record receives this evidence later, when the measurement repurifies the qubit. The global constant $c(\eta)=\inf DJ/\Gamma_m$ satisfies $K/2\le F_2\le c\le K$, where $F_2$ is again given by an integral; $c$ is known exactly at $\eta=1$ and in the limit $\eta\to0$. With any nonzero relaxation rate the unconstrained infimum of $DJ$ vanishes, so $c(\eta)$ is specific to the relaxation-free model.
\end{abstract}
\maketitle

\section{Introduction}
Continuous measurement turns the state of a quantum system into a classical signal, and quantum filtering turns that signal into a conditional state on which a controller can act \cite{Belavkin1992,Bouten2007,WisemanMilburn,JacobsSteck2006}. Measurement-based feedback built on this principle stabilizes and prepares states in cavity and circuit QED \cite{Sayrin2011,Vijay2012,Riste2012,CampagneIbarcq2013}, and individual quantum trajectories of superconducting qubits are routinely tracked \cite{Murch2013}. Much of the theory of such feedback asks how quickly, or how well, a target state is reached: rapid purification \cite{Jacobs2003,CombesJacobs2006,WisemanRalph2006,Jiang2019,Liu2026}, stabilization \cite{MirrahimiVanHandel2007}, and reset \cite{Geerlings2013,Magnard2018}.

Here we are interested in the record itself. A qubit monitored from the moment it is prepared in $|e\rangle$ writes evidence of that preparation into its output signal. Suppose a controller that knows the preparation and acts causally on the signal drives the qubit to $|g\rangle$. How much drive is needed so that the \emph{complete} record, including the part written before and during the drive, is hard to distinguish from the record of a qubit that was in $|g\rangle$ throughout?

The task is a hypothesis test. Under the reference hypothesis $H_0$, the qubit is prepared in $|g\rangle$ and nothing acts on it. Under $H_1$, it is prepared in $|e\rangle$, and a controller that knows this preparation applies a drive that depends causally on the record. An observer sees the measurement record $Y$ but has no access to the drive waveform or to the control hardware. The controller wants to reset the qubit while leaving the observer as little evidence of $H_1$ as possible. Evidence already in the record cannot be removed, so the controller can only shape what is written next, mainly by moving the qubit out of $|e\rangle$ quickly. We call this task \emph{record-indistinguishable reset}. The task should not be confused with Landauer erasure (we compute no thermodynamic quantity) or with the quantum eraser \cite{ScullyDruhl1982}, in which interference is restored by a later choice of measurement basis.

The figure of merit is the Kullback--Leibler (KL) divergence $D$ of the controlled record law from the reference law, which is the expected log-likelihood ratio that an observer accumulates when $H_1$ is true. The cost is the quadratic control effort $J=\E\int u^2dt$, the integrated squared Rabi frequency of a transverse drive. Faster driving lowers $D$ and raises $J$. Their product divided by the measurement rate $\Gamma_m$ is dimensionless, and its optimum depends only on the measurement efficiency $\eta$.

Two earlier lines of work meet in this problem. Passive hypothesis testing with continuous records \cite{Tsang2012,Molmer2015,bugu2026beyond} asks how well an observer can tell two dynamics apart; here a controller works against such an observer. State reset \cite{Geerlings2013,Magnard2018,Riste2012} asks only for the final state; here the cost is an integral over the whole history, so a slow reset that ends in $|g\rangle$ can still leave a highly informative record.

For relaxation rate $\Gamma_1=0$ and initial state $|e\rangle$ we obtain the following results. Open-loop drives leave $D=\infty$, regardless of their waveform or total effort (section~\ref{sec:openloop}). A potential of the Bloch length that does not depend on the control rewrites $D$ as an integrated rate containing an \emph{impurity price} (section~\ref{sec:potential}). Every admissible control obeys $DJ\ge\Gamma_mU(\eta,D/\kappa_0)$ (Theorem~\ref{thm:lower}). Finally, the infimum of $DJ/\Gamma_m$ over controls with $D\le\epsilon$ tends to $K(\eta)$ as $\epsilon\to0$, and feedback policies that we construct attain this limit (Theorem~\ref{thm:upper}).

The global constant $c(\eta)=\inf DJ/\Gamma_m$ lies between two functions of $\eta$ that differ by at most a factor of 2 and coincide at $\eta=1$ and as $\eta\to0$. We leave open whether $c(\eta)=K(\eta)$ for $0<\eta<1$. The product $DJ$ is a meaningful objective only without relaxation: for $\Gamma_1>0$ the undriven qubit already has finite $D$, and the unconstrained infimum of $DJ$ is zero (section~\ref{sec:relax}).

Section~\ref{sec:model} defines the model and the costs. Sections~\ref{sec:openloop}--\ref{sec:upper} state and explain the results, section~\ref{sec:mc} illustrates the construction by Monte Carlo simulation, and section~\ref{sec:discussion} discusses relaxation, related work, an illustrative parameter set and the limitations. Complete proofs are given in the Supplemental Material (SM).

\section{Model, hypotheses and costs}\label{sec:model}
\subsection{Dynamics}
A qubit is monitored in $\sigma_z$ with measurement rate $\Gamma_m$ and efficiency $\eta\in(0,1]$. It is driven by $H=\frac{u}{2}\sigma_y$, and $\Gamma_1=0$ except in section~\ref{sec:relax}. The conditional state obeys \cite{WisemanMilburn,JacobsSteck2006}
\begin{equation}\label{eq:sme}
d\rho=-i[H,\rho]dt+\Gamma_m(\sigma_z\rho\sigma_z-\rho)dt+\sqrt{\eta\Gamma_m}\,\mathcal H[\sigma_z]\rho\,dW,
\end{equation}
with $\mathcal H[\sigma_z]\rho=\sigma_z\rho+\rho\sigma_z-2\langle\sigma_z\rangle\rho$. The record is
\begin{equation}\label{eq:record}
dY=2\sqrt{\eta\Gamma_m}\,\langle\sigma_z\rangle\,dt+dW,
\end{equation}
where $W$ is the innovation Wiener process. Coherences decay at rate $2\Gamma_m$ on average; a fraction $\eta$ of the measurement is recorded, and the remaining fraction $1-\eta$ dephases the qubit without leaving a trace in $Y$. In Bloch coordinates the plane $y=0$ is invariant, and
\begin{equation}\label{eq:bloch}
\begin{aligned}
dx&=(uz-2\Gamma_mx)\,dt-2\sqrt{\eta\Gamma_m}\,zx\,dW,\\
dz&=-ux\,dt+2\sqrt{\eta\Gamma_m}\,(1-z^2)\,dW.
\end{aligned}
\end{equation}
The qubit starts in $|e\rangle$ ($z=1$).

\subsection{Admissible and operational controls}
An \emph{operational} policy is a drive $u_t=F_t(Y_{[0,t]})$ that is a nonanticipative functional of the record and for which the closed loop has a unique strong solution. The theorems are proved for a larger \emph{admissible} class: progressively measurable $u$ with $\E\int_0^tu^2<\infty$ for every $t$, on any filtration for which $W$ is a Brownian motion (SM, section~S1). The admissible class contains every operational policy, as well as controls that use randomness not contained in $Y$, so lower bounds proved for it hold for every operational policy. The construction of section~\ref{sec:upper} is operational.

\subsection{Record divergence}
We define
\begin{equation}\label{eq:D}
D=2\eta\Gamma_m\,\E\!\int_0^\infty(1+z_t)^2\,dt
\end{equation}
for every admissible control. Under $H_0$ the qubit stays in $|g\rangle$ and the record has the constant drift $-2\sqrt{\eta\Gamma_m}$. Let $\Pr_1$ and $\Pr_0$ be the laws of the record under $H_1$ and $H_0$. For an operational policy, $z_t$ is a functional of the record, and Girsanov's theorem \cite{LiptserShiryaev} gives $D=D_{\rm KL}(\Pr_1\|\Pr_0)$ (SM, Proposition~SA). Pinsker's inequality then implies that every test that uses the full record has error probability at least $\frac12(1-\sqrt{D/2})$ for equal priors \cite{CoverThomas}. For admissible controls that are not operational, $D$ is still well defined, but in general it is no longer a divergence of record laws.

We use the direction $D_{\rm KL}(\Pr_1\|\Pr_0)$ because it is the expected log-likelihood ratio in favor of $H_1$ when $H_1$ is true, which is the situation the controller is in. By Stein's lemma it is also the best exponent with which, over $n$ repeated independent runs, the probability of a false alarm on reference records can be made to decay while the probability of missing a controlled reset stays bounded \cite{CoverThomas}. It has the closed form~\eqref{eq:D} because the filter $z_t$ is evaluated on records generated under $H_1$. The reverse divergence $D_{\rm KL}(\Pr_0\|\Pr_1)$ evaluates the controller's filter on reference records. It is a different functional, generally with a different value, and we do not study it here. Pinsker's inequality bounds the total-variation distance by either divergence.

The divergence $D$ is computed on the infinite horizon and includes the part of the record written before and during the drive. It is finite only if the qubit is driven to $|g\rangle$ fast enough that $\E(1+z_t)^2$ is integrable.

\subsection{Control effort and scaling}
We use
\begin{equation}\label{eq:J}
J=\E\!\int_0^\infty u_t^2\,dt ,
\end{equation}
the integrated squared Rabi frequency, with units of s$^{-1}$ when $u$ is in rad/s. Its relation to microwave power, incident energy or photon number depends on how the drive is delivered and requires a device-specific calibration. For a resonant coherent drive through a single port with radiative coupling rate $\gamma_c$, input-output theory \cite{GardinerCollett1985} gives $u^2=4\gamma_c\Phi$, with $\Phi$ the incident photon flux, so $J=4\gamma_c\E N_{\rm in}$ (SM, section~S1C). Such a port also contributes $\gamma_c$ to the relaxation rate, so this conversion is compatible with $\Gamma_1=0$ only in the limit $\gamma_c\to0$, where the photon number for a given $J$ diverges. The quantity $J$ is a quadratic control-effort functional, not a calculation of thermodynamic work or heat.

In units of time $1/\Gamma_m$, $D$ is invariant and $J/\Gamma_m$ is dimensionless. The constant
\begin{equation}\label{eq:c}
c(\eta)=\inf\frac{DJ}{\Gamma_m},
\end{equation}
taken over admissible controls with $D,J<\infty$, therefore depends only on $\eta$. In the SM we set $\Gamma_m=1$.

\section{Feedback is necessary}\label{sec:openloop}
\begin{proposition}[Open-loop drives; SM, Proposition~SB]\label{prop:OL}
Let $\eta\in(0,1]$ and let $u$ be deterministic with $\int_0^tu^2ds<\infty$ for every $t$. Then $D=\infty$, whether the total effort $J=\int_0^\infty u^2dt$ is finite or not. If $J<\infty$, the unconditional excited population satisfies
\[
p(t)=\tfrac12(1+\E z_t)\ \ge\ \frac{4\Gamma_m}{J+4\Gamma_m}\quad\text{at every time }t,
\]
and the bound is approached by suitable drives as $t\to\infty$. A longitudinal drive $H\propto\sigma_z$, open loop or feedback, does not change $D$.
\end{proposition}

The proof uses only the averaged dynamics. For a deterministic drive, $(\bar x,\bar z)=(\E x,\E z)$ obeys $\dot{\bar x}=u\bar z-2\Gamma_m\bar x$ and $\dot{\bar z}=-u\bar x$, so $\frac{d}{dt}(\bar x^2+\bar z^2)=-4\Gamma_m\bar x^2$. The length of the mean Bloch vector therefore never increases, and it falls below 1 as soon as the drive has created any mean coherence. From then on $p$ is bounded away from zero, and since $\E(1+z)^2\ge4p^2$, $D=\infty$. Physically, coherence used for the rotation is dephased at rate $2\Gamma_m$, and without relaxation the resulting residual excitation never decays. For finite $J$, Cauchy--Schwarz bounds the rotation $1-\bar z$ by $(J\int\bar x^2)^{1/2}$, and this gives the bound on $p(t)$.

A controller that sees the record knows on which side of the equator the conditional state has landed and can correct it. The construction of section~\ref{sec:upper} has finite $D$ and $J$ for every $\eta>0$.

\section{The impurity price}\label{sec:potential}
The record divergence is an integral of the population signal $(1+z)^2$. The following identity re-expresses it through the Bloch length $r=\sqrt{x^2+z^2}$, which the drive cannot change.

\begin{proposition}[Potential identity; SM, Lemmas~S2 and S4]\label{prop:pot}
Let $h(r)=(1-r)-\ln\frac{2}{1+r}$, so that $h(1)=0$, $h(0)=\kappa_0=1-\ln 2$, and $h$ is concave and decreasing. For every control,
\begin{equation}\label{eq:pot}
\begin{split}
dh&=\big[G-2\eta\Gamma_m(1+z)^2\big]dt+dM,\\
G&=\Gamma_m\Big[\frac{2(1-\eta)x^2}{1+r}+\frac{2\eta(r+z)^2}{r}\Big],
\end{split}
\end{equation}
where $M$ is a martingale with bounded integrand. The control does not appear. If $D<\infty$, then $D=\E\int_0^\infty G\,dt$ and $\E h(r_t)\le D$ for all $t$.
\end{proposition}

The control drops out because $h$ depends only on $r$, and rotations preserve $r$. Equation~\eqref{eq:pot} identifies $G\ge0$ as the rate at which record divergence is charged, with $h$ as a balance that is settled when the state repurifies. Both terms of $G$ are easy to interpret. The term $2\eta\Gamma_m(r+z)^2/r$ plays the role of the recorded population signal; for a pure state it equals $2\eta\Gamma_m(1+z)^2$. The term $2(1-\eta)\Gamma_mx^2/(1+r)$ is the price of the impurity created when the unrecorded part of the measurement dephases the coherence $x$: the impurity $S=1-r^2$ is generated at rate $4(1-\eta)\Gamma_mx^2$ (SM, Lemma~S1; an equivalent It\^o equation for the impurity appears in reference~\cite{Liu2026}). Impurity created at time $t$ is invisible in the record at that time. It is paid for later, when the measurement repurifies the state near $|g\rangle$ and writes the outcome of that repurification into $Y$.

In polar coordinates $x=r\sin\theta$, $z=r\cos\theta$,
\begin{equation}\label{eq:Gpolar}
G=\Gamma_m\Big[\frac{2(1-\eta)r^2\sin^2\theta}{1+r}+2\eta r(1+\cos\theta)^2\Big]\le\Gamma_m\,g(\theta),
\end{equation}
\begin{equation}\label{eq:g}
g(\theta)=2\eta(1+\cos\theta)^2+(1-\eta)\sin^2\theta ,
\end{equation}
with equality at $r=1$. We call $g$ the \emph{angular cost density} [figure~\ref{fig:eta}(b)]. Near the excited state it is dominated by the population signal. Near the equator, where the rotation needs the most coherence, it contains the dephasing price $(1-\eta)\sin^2\theta$, which vanishes at $\eta=1$.

\begin{figure*}[t]
\includegraphics[width=\textwidth]{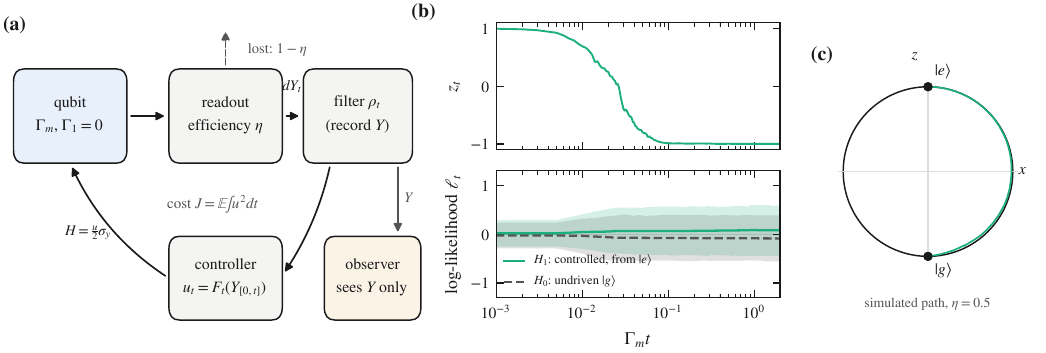}
\caption{Record-indistinguishable reset.
(a)~A qubit is measured continuously with efficiency $\eta$; a controller that knows the preparation acts on the record $Y$ with a transverse drive. The observer sees only $Y$. The cost is the quadratic control effort $J=\E\int u^2dt$.
(b)~Top: the conditional $z_t$ of one simulated trajectory at $\eta=0.5$, under the two-phase policy of section~\ref{sec:upper} with illustrative parameters ($\kappa=40$, $\beta=0.05$, $s=0.3$, $t_r=0.1$, time cap $T_{\max}=1/\Gamma_m$), not the parameter sequence of the proof. Bottom: running log-likelihood ratio $\ell_t=\ln(d\Pr_1/d\Pr_0)|_{[0,t]}$ for 400 simulated records under each hypothesis (lines: means; bands: 10--90\% quantiles; logarithmic time axis). Under $H_1$ the mean of $\ell_t$ is the accumulated divergence $D_t$, which saturates at $D$ once the qubit is reset.
(c)~The trajectory of (b) in the Bloch disk.}
\label{fig:schematic}
\end{figure*}

\section{Universal finite-budget bound}\label{sec:lower}
Define
\begin{equation}\label{eq:K}
K(\eta)=\Big(\int_0^\pi\sqrt{g(\theta)}\,d\theta\Big)^2,
\end{equation}
and, for $\lambda\in[0,1]$ and $\epsilon\ge0$,
\begin{equation}\label{eq:I}
\begin{split}
I_\lambda(\epsilon)=\int_0^\pi\!\!\Big\{&(1-\lambda)(1-\eta)\sin^2\varphi+2\eta\lambda(1+\cos\varphi)^2\\
&+2\eta(1-\lambda)\big[(1+\cos\varphi-\epsilon)_+\big]^2\Big\}^{1/2}d\varphi ,
\end{split}
\end{equation}
with $U(\eta,\epsilon)=\max_{\lambda\in[0,1]}I_\lambda(\epsilon)^2$ and $F_2(\eta)=U(\eta,2)$.

\begin{theorem}[Universal finite-budget bound]\label{thm:lower}
Let $\Gamma_1=0$, $\eta\in(0,1]$, and start in $|e\rangle$. Every admissible control with $D<\infty$ and $J<\infty$ satisfies
\begin{equation}\label{eq:lower}
DJ\ \ge\ \Gamma_m\,U\big(\eta,D/\kappa_0\big).
\end{equation}
$U(\eta,\cdot)$ is continuous and nonincreasing, with $U(\eta,0)=K(\eta)$ and $U(\eta,\epsilon)=F_2(\eta)$ for $\epsilon\ge2$. Hence $c(\eta)\ge F_2(\eta)$.
\end{theorem}

The bound holds for every admissible control and every $D$, without any high-effort assumption.

\subsection{Idea of the proof}
The proof (SM, section~S4) follows the mean $m(t)=\E z_t$ and combines three facts.

First, the drive moves the mean only through the coherence. From equation~\eqref{eq:bloch}, $\dot m=-\E[ux]$, so
\[
|\dot m|\le\sqrt{\E u^2}\,\sqrt{\E x^2}\qquad\text{and}\qquad \E x^2\le1-m^2 .
\]

Second, the record divergence can be written either as $2\eta\Gamma_m\int\E(1+z)^2$ or, by Proposition~\ref{prop:pot}, as $\int\E G$. Any convex combination of the two, with weight $\lambda$, also equals $D$. The second form contains the dephasing price $(1-\eta)\E x^2$, which ties $D$ to the coherence used for rotation.

Third, the potential identity gives $\E h(r_t)\le D$, so the state cannot become very impure: $\E r_t\ge1-D/\kappa_0$. This controls $(r+z)^2$ by $(1+m-D/\kappa_0)_+^2$ and is the origin of the second argument of $U$.

Combining these gives a density $\nu(m)$ with $\nu(m)|\dot m|\le\sqrt{\E u^2}\,\sqrt{\mathcal G_\lambda(t)}$, where $\int\mathcal G_\lambda dt\le D$. Cauchy--Schwarz in time gives $\int\nu(m)|\dot m|\,dt\le\sqrt{JD}$. The left side is at least the $\nu$-length of the path traversed by $m$ from $1$ to $-1$, which equals $I_\lambda(D/\kappa_0)$. The argument does not assume that the rotation is monotone. It uses only the absolute continuity of $m$, a last-exit time from a neighborhood of $m=1$, and a sequence of times along which $m\to-1$, which exists because $D<\infty$.

\subsection{Corollaries}
\begin{figure*}[t]
\includegraphics[width=\textwidth]{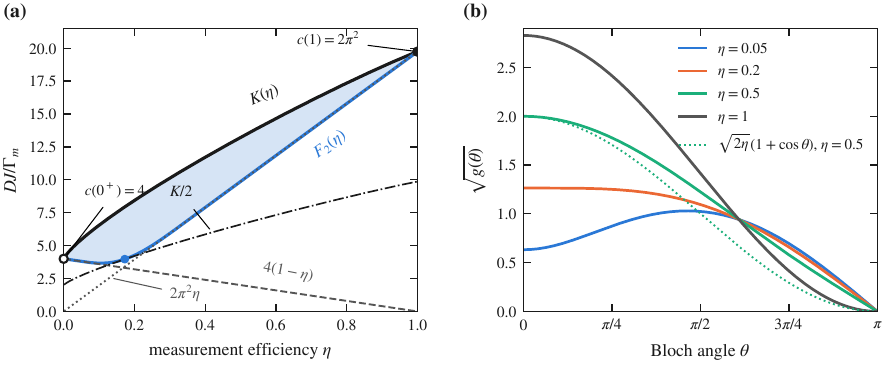}
\caption{Efficiency dependence. All curves are analytical.
(a)~The constant $c(\eta)=\inf DJ/\Gamma_m$ lies in the shaded band between the universal lower bound $F_2(\eta)$ (Theorem~\ref{thm:lower}) and the high-effort constant $K(\eta)$ (Theorem~\ref{thm:upper}). Also shown: the simpler lower bounds $2\pi^2\eta$ and $4(1-\eta)$, and $K/2$, which $F_2$ never falls below [equation~\eqref{eq:half}]; $F_2$ touches $K/2$ at $\eta\approx0.172$. The band closes at $\eta=1$, where $c=2\pi^2$, and as $\eta\to0$, where $c\to4$.
(b)~The angular cost density $\sqrt{g(\theta)}$ of equation~\eqref{eq:g}, whose integral is $\sqrt K$. Dotted: the recorded-signal part $\sqrt{2\eta}(1+\cos\theta)$ alone at $\eta=0.5$. The gap between the solid and dotted $\eta=0.5$ curves is the contribution of the impurity price; it is largest near and beyond the equator.}
\label{fig:eta}
\end{figure*}

Taking $\lambda=1$ in equation~\eqref{eq:lower} gives $DJ\ge2\pi^2\eta\Gamma_m$, which uses only the recorded signal. Taking $\lambda=0$ with $\epsilon\ge2$ gives $DJ\ge4(1-\eta)\Gamma_m$, which uses only the dephasing price. Taking $\lambda=\frac12$ with $\epsilon\ge2$ gives the integrand $\sqrt{g/2}$, hence
\begin{equation}\label{eq:half}
F_2(\eta)\ \ge\ \max\Big\{2\pi^2\eta,\ 4(1-\eta),\ \tfrac12K(\eta)\Big\}
\end{equation}
(SM, Lemma~SE). The bracket $F_2\le c\le K$ established below is therefore never wider than a factor of 2. Because $I_\lambda(2)$ is concave in $\lambda$, the equality $F_2=K/2$ holds exactly where $\lambda=\frac12$ is the optimal weight; numerically this happens at $\eta\approx0.172$ [figure~\ref{fig:eta}(a)].

As $D\to0$ the second argument of $U$ vanishes and the bound tightens to $K(\eta)$. A policy that beats $K$ by a fraction $\vartheta$ must therefore have $D\ge D_{\min}(\eta,\vartheta)$. For $\vartheta=1\%$, $D_{\min}=0.0094$, $0.0040$ and $0.0025$ at $\eta=0.05$, $0.2$ and $0.5$ (SM, Corollary~SC).

\section{Exact high-effort constant}\label{sec:upper}
\begin{theorem}[Exact high-effort constant]\label{thm:upper}
Let $\Gamma_1=0$, $\eta\in(0,1]$ be fixed, and start in $|e\rangle$. There is an explicit sequence of operational feedback policies with $D_n\to0$, $J_n\to\infty$ and
\begin{equation}\label{eq:upper}
\lim_{n\to\infty}\frac{D_nJ_n}{\Gamma_m}=K(\eta).
\end{equation}
Consequently $c(\eta)\le K(\eta)$, and
\begin{equation}\label{eq:main}
\lim_{\epsilon\downarrow0}\ \inf\Big\{\frac{DJ}{\Gamma_m}:\ \text{admissible},\ D\le\epsilon\Big\}=K(\eta).
\end{equation}
\end{theorem}

Theorem~\ref{thm:lower} gives the lower half of equation~\eqref{eq:main}, and the construction below gives the upper half.

\subsection{Heuristic argument}
If the rotation is fast compared with $\Gamma_m$, the state stays nearly pure and $\dot\theta\approx u$. Then by equation~\eqref{eq:Gpolar}
\[
D\approx\int\Gamma_mg(\theta)\,dt=\Gamma_m\int\frac{g(\theta)}{u}\,d\theta,\qquad J=\int u\,d\theta .
\]
Cauchy--Schwarz gives $DJ\ge\Gamma_m(\int\sqrt g\,d\theta)^2=\Gamma_mK$, with equality for $u\propto\sqrt{g(\theta)}$. The rotation should be slow where $g$ is small and fast where it is large. The constant $K$ is the squared length of the reset path in the metric $g\,d\theta^2$; thermodynamic length \cite{SalamonBerry1983,SivakCrooks2012,Scandi2019} and speed limits \cite{DeffnerCampbell2017,GarciaPintos2019} have the same structure.

This argument assumes a pure state and a deterministic angle. In fact the angle diffuses, the Bloch vector shrinks during the rotation, and the rotation ends in a mixed state near $|g\rangle$ from which the measurement can still send the qubit back toward $|e\rangle$. The more serious gap is that the dephasing part of $g$ is not written into the record during the rotation at all. It is stored as impurity, the balance $h$ in equation~\eqref{eq:pot}, and written into the record afterward, during repurification. The constant $K$ is therefore correct only if this deferred cost equals, to leading order, the charge $h$, uniformly over the random end state of the rotation, including trajectories that return toward $|e\rangle$ and must be caught by feedback. Proving this, with control of the stopping times and of the rare restarts, is the main work of the SM.

\subsection{Feedback construction}
In units $\Gamma_m=1$ the policy has two phases (SM, section~S5).
\begin{enumerate}
\item \emph{Rotation.} Apply $u=\kappa\sqrt{\tilde g(\theta)}$, with $\tilde g=g+\delta\sin^2\theta$ and $\theta=\operatorname{atan2}(x,z)$. Stop at $\tau'$, the first of three events: the angle reaches $\pi-\beta$ (good exit); the Bloch length drops to $\frac12$; or $t=1$.
\item \emph{Watch and restart.} Afterward set $u=0$. The conditional $z$ is then a bounded martingale that converges almost surely to $+1$ or $-1$. Whenever $z$ reaches a threshold $s$, apply a square pulse $u=\sqrt{1+(\pi/t_r)^2}$ of duration $t_r$. For a pulse that starts at $z=z_0$, the expected value of $z$ at its end is exactly $-z_0e^{-t_r}$, whatever the transverse coordinate; a pulse triggered at the threshold ends on average at $-se^{-t_r}$.
\end{enumerate}
The explicit sequence is
\[
\beta=\frac1\kappa,\qquad s=\kappa^{-1/8},\qquad t_r=s^3,\qquad \kappa=\kappa_n\to\infty,
\]
with $\delta=0$ for $\eta<1$ and $\delta=\kappa^{-1/10}$ for $\eta=1$. At $\eta=1$ the regularization is needed because $g$ vanishes to fourth order at $\theta=\pi$, and the hitting-time estimate of the rotation phase degenerates without it.

\subsection{Structure of the proof}
The proof (SM, sections~S5--S9) bounds $D$ and $J$ by closed-form quantities $\hat D$ and $\hat J$ (SM, Proposition~S20) and shows that $\kappa\hat D\to\sqrt K$ and $\hat J/\kappa\to\sqrt K$.

\emph{Rotation phase.} It\^o's formula for
\begin{gather*}
\Phi=\frac1\kappa\int_\theta^{\pi-\beta}\!\!\sqrt{\tilde g(\vartheta)}\,d\vartheta,\qquad \Psi=\int_0^\theta\frac{d\vartheta}{\sqrt{\tilde g(\vartheta)}},\\
\Phi_S=\frac1\kappa\int_\theta^{\pi-\beta}\!\frac{\sin^2\vartheta}{\sqrt{\tilde g(\vartheta)}}\,d\vartheta
\end{gather*}
gives three bounds, with $\Kt=(\int_0^\pi\sqrt{\tilde g}\,d\theta)^2$: $\E\int_0^{\tau'}\tilde g\le\sqrt{\Kt}/\kappa+O(\E\tau'/\kappa)$; a hitting time $\E\tau'=O(\ln\kappa/\kappa)$ with a Gaussian tail; and an impurity $\E S_{\tau'}=O(1/\kappa)$ with second moment $O(\ln^2\kappa/\kappa^2)$. The bad events (Bloch length below $\frac12$, or timeout) have probability $o(1/\kappa)$.

\emph{Deferred cost.} After the rotation, the remaining divergence is bounded by a value function $W(z)$ built from the exact off-phase quantities: the probability $(1+z)/(1+s)$ of returning to the threshold and the exact expected divergence before return. The bound rests on a Jensen inequality for the effect of each pulse, corrected for the kink of $W$ at $z=s$; on a geometric bound on the number of restarts; and on a supremum of the restart value over all pulse-start states, including pulses that start above the threshold because the previous pulse ended there. Near $|g\rangle$, $W(z)\le\Lambda(1+z)$ with $\Lambda\to\frac12=-h'(1)$. The deferred cost of the impurity left by the rotation therefore equals, to leading order, the potential $h$ that equation~\eqref{eq:pot} charges for it.

With $t_r=s^3$ the kink correction vanishes in the limit. With $t_r=s^2$ the bound would keep an $O(1)$ excess, although this need not reflect the actual cost of that policy. The gain $\kappa_1(\eta)$ above which the bounds apply grows as $\eta\to0$ and as $\eta\to1^-$ (SM, section~S9), and we have not shown that the convergence is uniform in $\eta$.

\subsection{The constant at the two ends of the efficiency range}
At $\eta=1$, $K(1)=2\pi^2=F_2(1)$. As $\eta\to0$, $K\to4$ and $F_2\ge4(1-\eta)\to4$. Hence
\begin{equation}\label{eq:ends}
c(1)=2\pi^2,\qquad \lim_{\eta\to0}c(\eta)=4
\end{equation}
(SM, Corollary~SC). For intermediate $\eta$ the bracket $F_2\le c\le K$ is open, and by equation~\eqref{eq:half} its ratio never exceeds 2. SM section~S11 reports a one-dimensional numerical solution of the Bellman equation at $\eta=1$, obtained with a Markov-chain approximation \cite{KushnerDupuis}. It agrees with the proven value $c(1)=2\pi^2$, although we have not proved that the scheme converges.

\begin{figure*}[t]
\includegraphics[width=\textwidth]{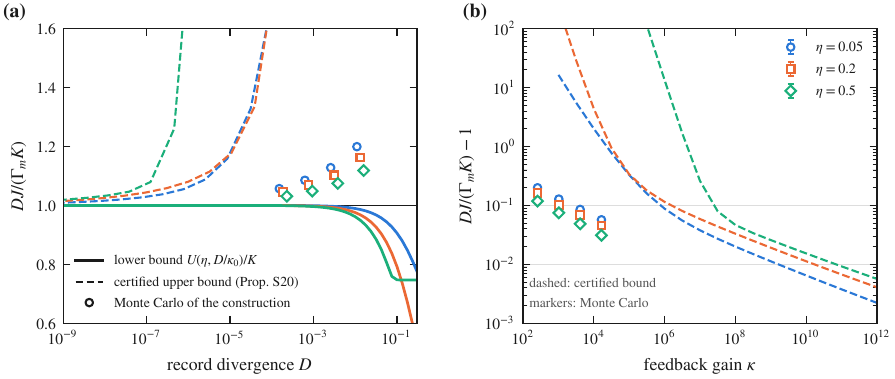}
\caption{The high-effort theorem.
(a)~Solid: the universal lower bound $U(\eta,D/\kappa_0)/K$ of Theorem~\ref{thm:lower}; no admissible control lies below its curve. Dashed: the certified closed-form upper bound $(\hat D,\hat D\hat J/K)$ for the construction (SM, Proposition~S20), traced as $\kappa$ increases. Markers: Monte Carlo of the same construction with $\kappa=2^8,2^{10},2^{12},2^{14}$ and the proof's parameter sequence, using the conditional estimator of section~\ref{sec:mc} ($3.2\times10^4$ trajectories each). Statistical error bars (one standard error) are smaller than the markers; the spread over step sizes is at most $0.3\%$ of $K$. Both bounds converge to $1$ as $D\to0$.
(b)~Excess $DJ/(\Gamma_mK)-1$ against the dimensionless gain $\kappa$. Dashed: the certified bound, which comes within $10\%$ of $K$ for $\kappa\gtrsim0.8$--$23\times10^6$ and within $1\%$ for $\kappa\gtrsim1.6$--$69\times10^9$ ($\eta=0.05$ to $0.5$). Markers: Monte Carlo. These gains are dimensionless; section~\ref{sec:exp} relates them to Rabi frequencies.}
\label{fig:high}
\end{figure*}

\section{Monte Carlo illustration of the construction}\label{sec:mc}
The numerical results in this section are not used in any proof. They show how the construction behaves at accessible gains, where the certified bound is far from tight (figure~\ref{fig:high}). Methods are in SM section~S11.

The certified bound is conservative: it comes within $10\%$ of $K$ only for $\kappa\approx0.8$--$23\times10^{6}$, and within $1\%$ only for $\kappa\approx1.6$--$69\times10^{9}$ ($\eta=0.05$ to $0.5$). The simulated policy is much closer to $K$. At $\kappa=2^{14}$ the simulation gives
\[
\frac{DJ}{\Gamma_mK}=1.0570(14),\ \ 1.0454(10),\ \ 1.0311(5)
\]
for $\eta=0.05$, $0.2$ and $0.5$ (statistical errors in the last digits), at $D\approx(1.5$--$2.3)\times10^{-4}$. Every estimate lies above the lower bound $\Gamma_mU(\eta,D/\kappa_0)$, as it must. No exits through $r=\frac12$ and no timeouts occurred.

Restarts are rare and costly, and this makes a direct simulation of every trajectory inefficient. In direct simulation with $6.4\times10^4$ trajectories per point, the trajectories that contain a restart carry $3$--$16\%$ of the mean divergence but almost all of its variance, and at $\kappa=2^{14}$ only 5 to 13 trajectories restart. The discretely monitored threshold also biases the off-phase divergence, by an amount of order $\sqrt{dt}$ (SM section~S11). We therefore use a conditional estimator. Phase 1 and every pulse are simulated, but the off phase is replaced by its exact solution (SM, Lemma~S15): the expected divergence $f(z)$ before the next pulse, the probability $(1+z)/(1+s)$ that a next pulse occurs, and a sample of the state in which it starts, drawn from its exact conditional law. The rare restarts thus enter through their exact probabilities rather than through sampled events, and the threshold error disappears.

The conditional estimates are stable under refinement of the time step. For step factors 1, $\frac12$ and $\frac14$ at all twelve points, and down to $\frac1{16}$ at one point, the values agree within their statistical errors ($\chi^2=20.8$ for 28 degrees of freedom), and the largest spread at any point is $0.3\%$ of $K$. We take this as empirical evidence of convergence; it does not bound the discretization error. Direct simulations with the original code agree with the conditional estimates within $1.4$ of their standard errors at all twelve points, and an independently written integrator agrees within one standard error at seven of eight points and differs by $2.2$ at $(\eta,\kappa)=(0.5,2^{10})$ (SM table~S2). Direct simulation at $(\eta,\kappa)=(0.5,2^{10})$ with several step sizes and seed lists scatters by up to three of its quoted standard errors, without a monotone trend. We found no algorithmic cause in the phase-1, threshold or pulse handling and attribute the scatter to fluctuations of the small number of restarts, superposed on the threshold error (SM section~S11).

\section{Discussion}\label{sec:discussion}
\subsection{Interpretation of the constant}
The angular cost density $g$ in equation~\eqref{eq:g} splits the cost of a record-indistinguishable reset into two parts. The recorded population signal, $2\eta(1+\cos\theta)^2$, is the evidence written while the qubit is still excited; on its own it gives $2\pi^2\eta$. The dephasing price, $(1-\eta)\sin^2\theta$, arises because rotating a qubit requires coherence and the unrecorded part of the measurement turns that coherence into impurity, which the record registers only later, when the measurement repurifies the qubit. At low efficiency the second part dominates and $c\to4$. At unit efficiency it vanishes, and the cost is set by how long the record keeps showing an excited qubit. Lowering $\eta$ lowers $K$, since a weaker record carries less evidence, but by less than it lowers the recorded signal: $K(\eta)/(2\pi^2\eta)$ equals 1 at $\eta=1$, increases monotonically as $\eta$ decreases, and diverges as $\eta\to0$. Relative to the information that is actually recorded, the unrecorded part of the measurement makes the reset more expensive.

\subsection{Relaxation}\label{sec:relax}
The product objective is specific to $\Gamma_1=0$. Add amplitude damping at rate $\Gamma_1>0$, with the emitted field not monitored, and apply no drive. The unconditional excited population is $p(t)=e^{-\Gamma_1t}$. Since $0\le1+z\le2$, $\E(1+z_t)^2\le2\E(1+z_t)=4p(t)$, and
\begin{equation}\label{eq:relax}
D\big|_{u=0}\ \le\ 8\eta\,\frac{\Gamma_m}{\Gamma_1}<\infty,\qquad J=0
\end{equation}
(SM, Proposition~SF). The unconstrained infimum of $DJ$ is therefore zero for every $\Gamma_1>0$, and the constant $c(\eta)$, its bracket and the question whether $c=K$ exist only in the relaxation-free model.

The small-$D$ question remains meaningful. For $\epsilon\ll8\eta\Gamma_m/\Gamma_1$ the controller must still remove the excitation much faster than relaxation does, and one can ask for $\lim_{\epsilon\to0}\inf\{DJ/\Gamma_m: D\le\epsilon\}$. We do not answer this question for $\Gamma_1>0$: the potential identity, the lower bound and the restart analysis all change when relaxation adds unrecorded jumps, so Theorem~\ref{thm:upper} does not carry over unchanged.

\subsection{Relation to existing theory}
The mathematical tools are standard: quantum filtering, It\^o calculus, the Cauchy--Schwarz inequality and stopping-time estimates \cite{RevuzYor}. The squared-length form of $K$ is the template of thermodynamic length \cite{SalamonBerry1983,SivakCrooks2012}, including its open-quantum-system version \cite{Scandi2019}, and of quantum speed limits \cite{DeffnerCampbell2017}, which have also been analyzed along continuously measured trajectories \cite{GarciaPintos2019}. Those works bound dissipation or evolution time along a prescribed path. In our problem the metric itself contains a cost that the record registers only later, and the result is an exact asymptotic constant for a feedback-controlled mixed-state diffusion.

The link between a KL divergence of path laws and a quadratic drift cost is the basis of KL and path-integral control \cite{Todorov2009,Kappen2012} and of Schr\"odinger-bridge formulations, which have recently been extended to continuously measured quantum systems \cite{JordanOhzeki2026}. In those settings the controller pays for moving its own path law away from a passive reference. Here the controller pays a quadratic effort to keep the observer's record law close to a reference, so the KL divergence plays the opposite role: it is the objective, and the effort is the cost. Optimal steering of a monitored qubit, with a measurement-derived cost, has been studied for Zeno dragging \cite{Lewalle2024}, where the control target is the state itself.

Feedback accelerates purification \cite{Jacobs2003,CombesJacobs2006,WisemanRalph2006,Jiang2019,Liu2026}, and preparing a state by measurement requires at least a final feedback step \cite{CombesWisemanScott2010}; Proposition~\ref{prop:OL} quantifies a related separation for record-indistinguishable reset. Thermodynamic analyses of continuous feedback \cite{SagawaUeda2008,Prech2025,GarciaNila2026} are primarily concerned with work, heat and entropy production; the distinguishability of the output record is not their focus. The game structure, acting while keeping an observer's divergence small, is that of covert communication and covert sensing \cite{Bash2015,Bash2017}, where square-root laws limit the information that can be transferred under a divergence constraint. We are not aware of earlier results on the drive cost of record indistinguishability for a monitored qubit.

\subsection{Illustrative circuit-QED parameters}\label{sec:exp}
The model is the standard diffusive-readout model of a superconducting qubit \cite{Gambetta2008,Murch2013,Blais2021}. Throughout, $\kappa$ is the dimensionless feedback gain; it is unrelated to the resonator linewidth that is often denoted by the same symbol. To relate it to physical drives, take $\Gamma_m/2\pi=0.1$~MHz as an illustrative value, so $1/\Gamma_m\approx1.6~\mu$s, comparable to the microsecond integration windows of trajectory experiments \cite{Murch2013}. The actual $\Gamma_m$ depends on the dispersive shift, the readout photon number and the resonator linewidth. A strong readout drive can itself leave population in higher transmon levels \cite{bugu2026_LORC_persistent}, which the two-level model~\eqref{eq:sme} does not describe.

In the rotation phase $u=\kappa\Gamma_m\sqrt{g(\theta)}$. For $\eta\ge1/5$ the maximum of $g$ is $8\eta$, at $\theta=0$; for $\eta<1/5$ it is $1+\eta+4\eta^2/(1-3\eta)$, at $\cos\theta=2\eta/(1-3\eta)$. At $\eta=0.5$ the peak Rabi frequency is $u/2\pi=2\kappa\times0.1$~MHz, so $\kappa=2^{10}$ (where the simulation gives $DJ/K\approx1.075$) requires about $205$~MHz, and $\kappa=2^{12}$ about $820$~MHz. Transmon anharmonicities are $-E_C/h$ with $E_C/h\sim100$--$400$~MHz \cite{Blais2021}. Rabi frequencies of this size populate the second excited state and invalidate the two-level and dispersive-readout approximations on which equation~\eqref{eq:sme} rests; with these parameters the two-level description is plausible only for $\kappa$ of order $10^2$ or below. We have not simulated that range. At the smallest simulated gain, $\kappa=2^8$ (peak Rabi frequency about $51$~MHz at $\eta=0.5$), $DJ/K$ is already $1.12$--$1.20$, and in each series of figure~\ref{fig:high} the excess increases monotonically as $\kappa$ decreases.

The rotation phase lasts about $18$~ns at $\kappa=2^{10}$ and $5$~ns at $\kappa=2^{12}$ ($\eta=0.5$), whereas a low-latency FPGA feedback loop has a latency of about $110$~ns \cite{Salathe2018}, so state feedback during the rotation is not available. Because the rotation is nearly deterministic (impurity $O(1/\kappa)$), a precomputed pulse might replace it, with feedback used only in the restart phase, which acts on the scale $1/\Gamma_m$ with pulses of length $t_r\approx120$~ns at $\kappa=2^{10}$. Proposition~\ref{prop:OL} does not exclude this variant, because the restart phase remains closed loop, but we have not analyzed it. Even there the $110$~ns latency is not negligible: at $\kappa=2^{10}$ and $\eta=0.5$ it is comparable to the time in which $z$ diffuses across a distance of order $s$ near the threshold.

With an illustrative $T_1=100~\mu$s, $\Gamma_1/\Gamma_m\approx0.016$, and the passive bound~\eqref{eq:relax} is $D\lesssim250$ at $\eta=0.5$, far above the controlled values $D\sim10^{-3}$. Relaxation is negligible during the rotation, which lasts about $10^{-2}/\Gamma_m$. In the post phase, in which the deferred cost is paid, the equation $dz=2\sqrt{\eta\Gamma_m}(1-z^2)\,dW$ depends on time only through $\eta\Gamma_mt$. The deferred cost is therefore paid on the time scale $1/(\eta\Gamma_m)$: in simulations of the construction, the divergence-weighted mean time after the rotation is $2$--$3/\Gamma_m$ at $\eta=0.5$ and about $20/\Gamma_m$ at $\eta=0.05$. Relaxation thus perturbs the post phase at relative order $\Gamma_1/(\eta\Gamma_m)$. This is about $0.03$ at $\eta=0.5$, but about $0.3$ at $\eta=0.05$, where the perturbation is not small and the $\Gamma_1=0$ results should not be applied quantitatively. Its effect on $D$ is not quantified here.

Lowering $\Gamma_m$ allows larger $\kappa$ at a fixed Rabi frequency but makes relaxation relatively stronger. With present transmon parameters the asymptotic regime of Theorem~\ref{thm:upper} is out of reach, and the simulations of section~\ref{sec:mc} describe the finite-gain behavior instead. Comparing measured $DJ$ with $K(\eta)$ would also require a calibration of $J$ and many repetitions to estimate a small $D$.

\subsection{Scope and an open question}
Besides $\Gamma_1=0$ (section~\ref{sec:relax}), the theorems assume that the initial state is $|e\rangle$ and that the observer sees only $Y$; an observer with access to the drive line could distinguish the hypotheses trivially. Theorem~\ref{thm:upper} holds for each fixed $\eta$, with gain thresholds that grow as $\eta\to0$ or $\eta\to1^-$. Feedback is assumed instantaneous and of unlimited bandwidth, and $J$ is a quadratic control effort whose conversion to power or photon number depends on the implementation.

Since $c(\eta)=\inf_\mu V_\mu^2/(4\mu\Gamma_m)$ with $V_\mu=\inf(D+\mu J)$ (SM, Lemma~SD), a verification argument for the Bellman equation at finite $\mu$ would decide whether $c(\eta)=K(\eta)$ for $0<\eta<1$, that is, whether any finite-effort policy beats the high-effort limit. We present no numerical evidence on this question.

\section{Conclusions}
Resetting a monitored qubit while keeping its record close to that of a ground-state qubit requires feedback, because every open-loop drive leaves the record divergence infinite. With feedback, the product of the record divergence and the quadratic control effort is bounded below, for every admissible control, by a function $\Gamma_mU(\eta,D/\kappa_0)$ of the efficiency and of the divergence itself. For $\Gamma_1=0$, an initial excited state and fixed $\eta\in(0,1]$, the infimum of this product over controls with vanishing divergence equals $\Gamma_mK(\eta)$. This is the squared length of the reset path in a metric that adds the recorded population signal to the price of the impurity created by unrecorded dephasing, a price the record collects only after the rotation. Two-phase feedback policies attain it. The global constant is known exactly at unit efficiency and in the zero-efficiency limit, and between these limits it is bracketed to within a factor of 2. It is specific to the relaxation-free model, because any relaxation makes the unconstrained product infimum vanish.

\bibliography{pcqe}
\end{document}